\documentclass[aps,twocolumn,superscriptaddress]{revtex4-2}
\usepackage{amsfonts}
\usepackage{dcolumn}
\usepackage{bm}
\usepackage{tikz}
\usepackage[colorlinks=true,citecolor=blue,urlcolor=blue]{hyperref}
\usepackage{amsmath,amsfonts,amssymb,times,natbib}
\usepackage[standard]{ntheorem}

\begin{document}

\title{Simulating topological materials with photonic synthetic dimensions in cavities}

\author{Mu Yang}
\affiliation{CAS Key Laboratory of Quantum Information, University of Science and Technology of China, Hefei 230026, China}
\affiliation{CAS Center For Excellence in Quantum Information and Quantum Physics, University of Science and Technology of China, Hefei 230026, China}

\author{Jin-Shi Xu}\email{jsxu@ustc.edu.cn}
\affiliation{CAS Key Laboratory of Quantum Information, University of Science and Technology of China, Hefei 230026, China}
\affiliation{CAS Center For Excellence in Quantum Information and Quantum Physics, University of Science and Technology of China, Hefei 230026, China}
\affiliation{Hefei National Laboratory, University of Science and Technology of China, Hefei 230088, China}

\author{Chuan-Feng Li}
\email{cfli@ustc.edu.cn}
\affiliation{CAS Key Laboratory of Quantum Information, University of Science and Technology of China, Hefei 230026, China}
\affiliation{CAS Center For Excellence in Quantum Information and Quantum Physics, University of Science and Technology of China, Hefei 230026, China}
\affiliation{Hefei National Laboratory, University of Science and Technology of China, Hefei 230088, China}

\author{Guang-Can Guo}
\affiliation{CAS Key Laboratory of Quantum Information, University of Science and Technology of China, Hefei 230026, China}
\affiliation{CAS Center For Excellence in Quantum Information and Quantum Physics, University of Science and Technology of China, Hefei 230026, China}
\affiliation{Hefei National Laboratory, University of Science and Technology of China, Hefei 230088, China}

\begin{abstract}
Photons play essential roles in fundamental physics and practical technologies. They have become one of the attractive informaiton carriers for quantum computation and quantum simulation. Recently, various photonic degrees of freedom supported by optical resonant cavities form photonic synthetic dimensions, which contribute to all-optical platforms for simulating novel topological materials. The photonic discrete or continuous degrees of freedom are mapped to the lattices or momenta of the simulated topological matter, and the couplings between optical modes are equivalent to the interactions among quasi-particles. Mature optical modulations enable flexible engineering of the simulated Hamiltonian. Meanwhile, the resonant detection methods provide direct approaches to obtaining the corresponding energy band structures, particle distributions and dynamical evolutions. In this Review, we give an overview of the synthetic dimensions in optical cavities, including frequency, orbital angular momentum, time-multiplexed lattice, and independent parameters. Abundant higher-dimensional topological models have been demonstrated in lower dimensional synthetic systems. We further discuss the potential development of photonic synthetic dimensions in the future.
\end{abstract}

\date{\today}

\maketitle
Photonic synthetic dimensions, first proposed in 2015~\cite{luo2015quantum}, greatly enrich the field of topological photonics~\cite{yuan2018synthetic,ozawa2019topological,price2022roadmap}. The basic idea for photonic synthetic dimensions is to construct a series of lattice or momenta by using photonic intrinsic degrees of freedom to study distinguished phenomena in condensed matter physics. The optical modes are regarded as quasi-particles, while the transition among optical modes is equivalent to the interaction among them. By introducing mature optical elements to engineer the coupling, we can conveniently regulate and control the Hamiltonian of the simulated topological materials. Moreover, various unique optical detection methods have been developed based on the photonic synthetic dimension, from which we can directly visualize the pivotal information of the corresponding topological phases. Compared with conventional photonic crystals~\cite{mukherjee2020observation,maczewsky2020nonlinearity,zhou2018observation,jin2019topologically,yin2020observation,stutzer2018photonic,rechtsman2013photonic,weimann2017topologically,bandres2018topological,hafezi2013imaging,lu2016topological,mittal2016measurement} based on photon propagation and interference in real space, it is possible to directly investigate the dynamical properties of the simulated system in synthetic space. 

The simulated physical dimensions can be significantly increased in synthetic space, which is independent of the real space.  We can simulate a ($D+d$)-dimensional ($D+d>3$) physical system in a $D$-dimensional real system with $d$ synthetic dimensions. Moreover, systems based on synthetic dimensions usually have the capability of dynamic control, which benefits the study of concrete topological phenomena. In addition, photonic synthetic dimensions provide new insight into the device design~\cite{luo2017synthetic}. 
 
Photonic degrees of freedom have been used in investigating topological physics with periodically stacked structures, however, these systems are limited dimensional scaling~\cite{xiao2017observation,xiao2020non,zhan2017detecting,wang2019simulating,cardano2017detection,cardano2015quantum,wang2018dynamic,xu2018measuring}. Optical resonant cavities have been employed to construct synthetic dimensions, which can trap photons in a zero dimensional space. The photonic degrees of freedom that can be used to construct synthetic dimensions in cavities include frequency~\cite{yuan2021synthetic,leykam2020topological}, spin and orbital angular momentum (SAM and OAM)~\cite{yang2022topological,luo2015quantum}, time-multiplexed lattice~\cite{regensburger2012parity,adiyatullin2022multi}, independent parameters~\cite{miri2019exceptional} and so on.  The Hamiltonian can be conveniently engineered by introducing optical modulators into the cavity. The principles of mode self-reproduction in the cavity and coherent output provide ingenious optical detection methods, from which the energies structure, particle distributions and dynamics of the simulated system can be directly extracted.

In this Review, we overview the realization of synthetic dimensions in cavities, including frequency, OAM,  time-multiplexed lattice, and independent parameters. Different synthetic dimensional platforms constructed by fiber loops, degenerate optical cavities, and microcavties are introduced. We provide simplified setups for constructing synthetic dimensions and discuss the detection methods of energy band spectroscopy, particle distributions and dynamical behavior in various setups. At the end of each section, we review the recent progresses and applications of corresponding photonic synthetic dimensions. Finally, we prospect the potential developments of interphoton interactions along synthetic dimensions in studying many-body physics and high-dimensional topological materials .\\

\begin{figure*}[t!]
	\begin{center}
		\includegraphics[width=1.9 \columnwidth]{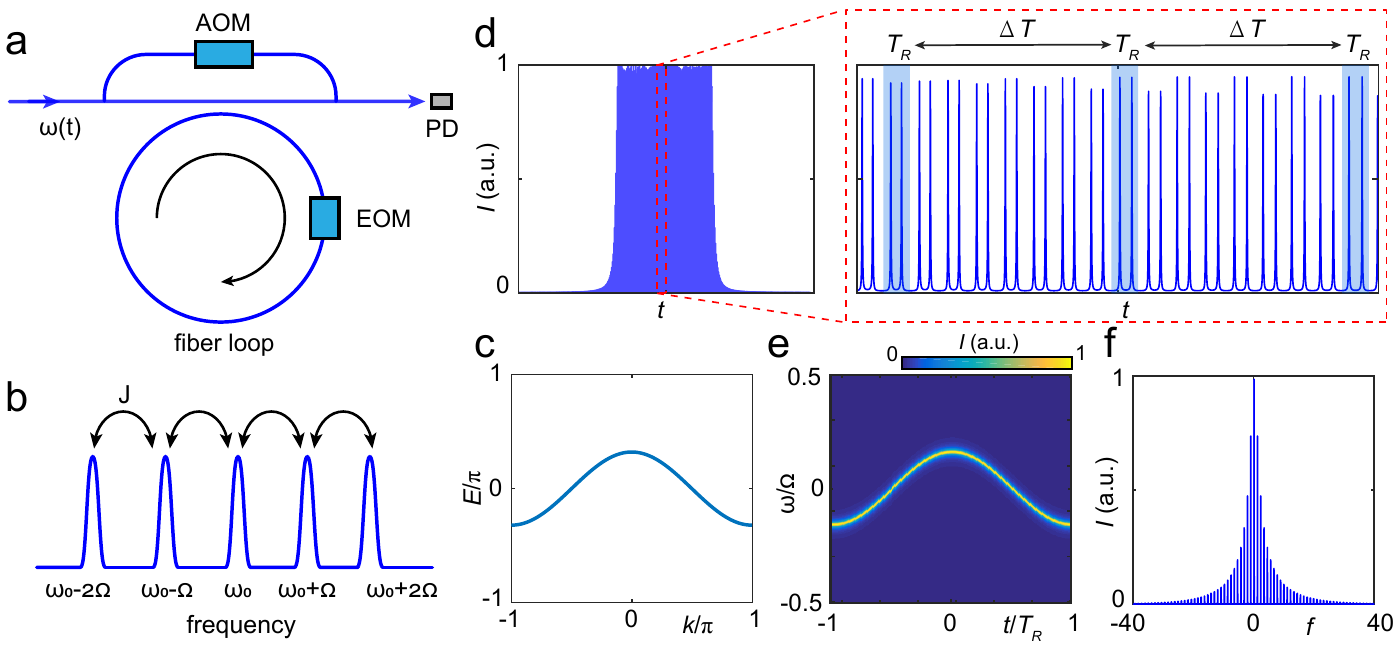}
		\caption{ \textbf{Synthetic frequency dimension.} (a) Simplified setup containing a fiber loop with resonant frequency $\Omega$. AOM: acoustic-optic modulato; EOM: electro-optic modulator; PD: photoelectric detector.  (b) The coupling among frequency lattice with coupling strength $J$. (c) The calculated energy band of the system. (d) The transmitted intensity spectrum vs time.  (e) The energy band extracted by time slicing. (f) The frequency distribution as $E_{f}=0$. 
		}
		\label{f}
	\end{center}
\end{figure*}


\noindent\textbf{Frequency}\\
The key to synthesize a photonic lattice is to construct a series of optical modes labelled by discrete integers. We firstly discuss the frequency degree of freedom, which can clearly illustrate the process in constructing photonic syntehtic space. For the synthetic frequency dimension, the frequency modes $|n\rangle$ in a fiber loop resonator (as shown in Fig. \ref{f}a) can be discrete as $|n\rangle=Ae^{in\Omega t}$ ($n\in\mathbb{N}$), where $A$ is the amplitude, $\Omega$ represents the resonant frequency of the cavity and $t$ is the evolution time. 
As we introduce an electro-optic modulator (EOM), loaded a radio frequency (RF) with frequency $\Omega$, into the resonator, the optical modes $|n\rangle$ will become $2J\cos\Omega t|n\rangle=JA(e^{i(n+1)\Omega t}+e^{i(n-1)\Omega t})=J(|n+1\rangle+|n-1\rangle)$. Thus the photons with the frequency of $n\Omega$ and $(n\pm1)\Omega$ will be coupled (see the hoping model in Fig. \ref{f}b) with coupling strength $J$, which can be controlled by the voltage of the RF~\cite{yuan2016bloch}. It is worthy to note the coupling strength $J$ among the frequency modes is usually set small so that the long-range connections introduced by the high harmonic modulation of EOM are weak enough to be ignored. The system can be well described through the tight-binding method, and the Hamiltonian of the system can be written as
\begin{equation}
\begin{aligned}
H_{f}=\sum_{n}(J\hat{a}_{n+1}^{\dag}\hat{a}_{n}+\mathrm{h.c.}),
\end{aligned}
\end{equation}
where $\hat{a}^{\dag} (\hat{a})$ represents the creation (annihilation) operator of the frequency modes. $\hat{a}_{n}$ satisfies the Bloch condition, we can transform it into momentum ($k$) space with $\hat{a}_{k}=\sum_{n}\hat{a}_{n}e^{-ikn}$. The Hamiltonian can be written as $H_{f}(k)=\sum_{k}\hat{a}_{k}^{\dag}\hat{a}_{k}E_{f}(k)$, where $E_{f}(k)=2J\cos k$ is the eigenenergy of this synthetic frequency lattice. The calculated energy band as $J=0.1$ is shown in Fig. \ref{f}c. 

As we input photons with frequency $\omega$ into the fiber loop in Fig. \ref{f}a, the input and output relation of the cavity can be described by the Heisenberg-Langevin equation~\cite{scully1999quantum}, denoted as
\begin{equation}
\begin{aligned}
&\frac{d\hat{a}_{n}(t)}{dt}=-i[\hat{a}_{n}, H_{f}]-\frac{\gamma}{2}\hat{a}_{n}(t)-\sqrt{\gamma}\hat{d}_{in, n}(t),\\
&\hat{d}_{out, n}(t)=\sqrt{\gamma}\hat{a}_{n}(t),
\end{aligned}\label{L}
\end{equation}
where $\gamma$ is the loss rate of the cavity. $\hat{d}_{in, n}(t)$ and $\hat{d}_{out, n}(t)$ represent the input and output operators, respectively. Applying Fourier transform to both sides of Eq.~\ref{L}, the solution of the output filed $\hat{d}_{out, n^{'}}(\omega)=\frac{1}{\sqrt{2\pi}}\int_{-\infty}^{\infty}d\omega e^{-i\omega t}\hat{d}_{out, n^{'}}(t)$ is $\hat{d}_{out, n^{'}}(\omega)=T^{n^{'}}_{n}\hat{d}_{in, n}(\omega)$. $T^{n^{'}}_{n}$ represents the transmitted coefficient defined as
\begin{equation}
\begin{aligned}
T^{n^{'}}_{n}=\langle n^{'}|\frac{\gamma}{\omega-H_{f}+i\gamma/2}|n\rangle.
\end{aligned}\label{T}
\end{equation}
The transmitted beam intensity containing all frequency modes is $I_{n}=\sum_{n^{'}}|T^{n^{'}}_{n}|^2\propto\gamma^2/[(\omega-E_{f})^2+\gamma^2/4)]$, where $E_{f}$ is the eigenenergy of the system.  Only the frequency of the driving light approaching the eigenenergy, the cavity has a maximum transmission. 

\begin{figure*}[t!]
	\begin{center}
		\includegraphics[width=1.9 \columnwidth]{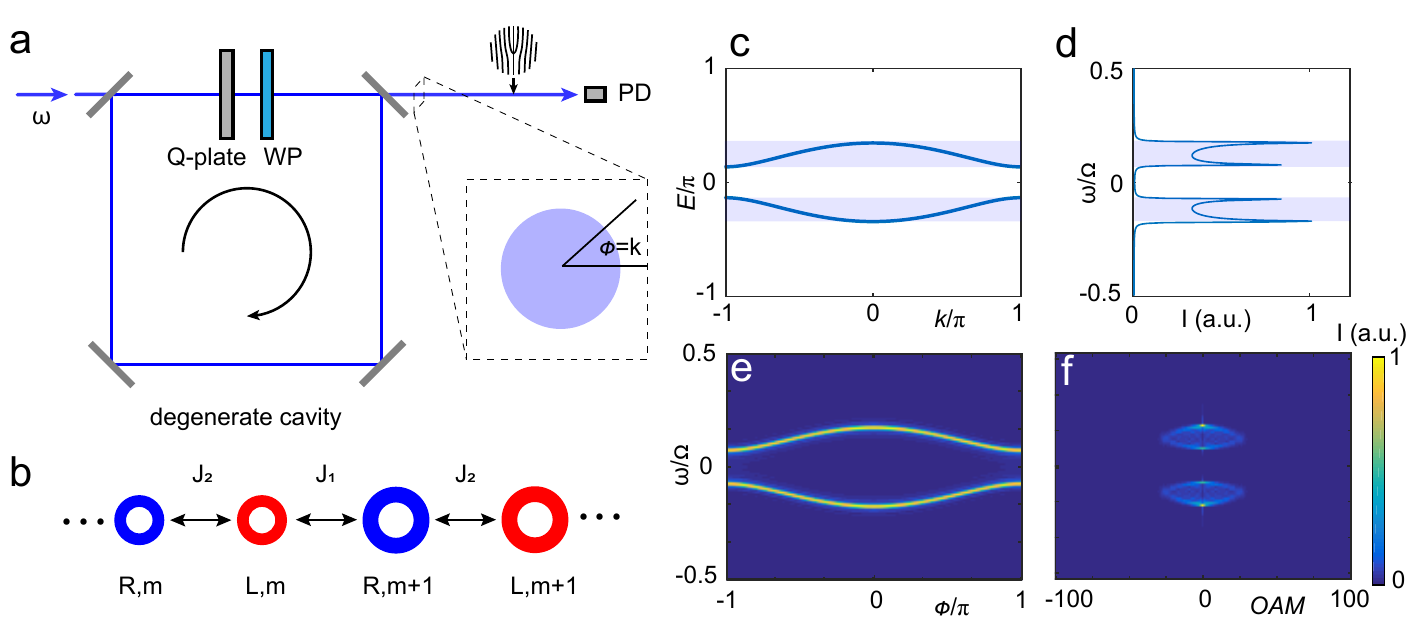}
		\caption{ \textbf{Synthetic OAM dimension.} (a) Simplified setup containing a degenerate cavity. WP: wave plate; PD: photoelectric detector. (b) The coupling among different spin and orbital angular momentum (SAM and OAM) with coupling strength $J_{1}$ or $J_{2}$. R: right-circular polarization; L: left-circular polarization;  (c) The calculated energy bands. (d) The transmitted intensity spectrum vs input frequency $\omega$. (e) The energy bands by projecting the output photons along the different wavefront azimuthal angle $\phi$. (f) The OAM distribution of the output photons projected by forked gratings. 
		}
		\label{OAM}
	\end{center}
\end{figure*}

The Bloch state along frequency lattice is $|k\rangle=\sum_{n}e^{-ink}|n\rangle$ and the Bloch state can be regarded as $|k\rangle=A\sum_{n}e^{-in(k-\Omega t)}=A\delta(k,\Omega t)$. The time $t$ corresponds to the momentum $k/\Omega$, and $T_{R}=2\pi/\Omega$ corresponds to the first Brillouin zone. 
By linearly sweeping the frequency $\omega$ of the input photons, the output intensity spectrum vs time can be detected via an oscilloscope, which is shown in Fig. \ref{f}d. The energy band can be extracted via time slicing~\cite{li2021dynamic,dutt2020single} as schemed in the red box in Fig. \ref{f}d. The intensity spectrum within a time slice of $T_{R}$ at each $\omega(t_{0}+l\Delta T)$ ($l\in\mathbb{N}^{+}$) are sliced for reconstructing the energy band spectrum, where $t_{0}$ is the starting time and $\Delta T$ represents the interval of the time slices. The band spectrum of $J=0.1$ and $\gamma=0.05$ is shown in Fig. \ref{f}e by sorting the intensity spectra of different time slices column by column. Moreover, the probability distributions of frequency modes can be read out via heterodyne measurement between input and output photons. The readout energy levels can be adjusted by frequency shifting via an acoustic-optic modulator (AOM) as shown in Fig. \ref{f}a. The frequency mode distribution as $E_{f}=0$ when $J=0.1$ and $\gamma=0.05$ is shown in Fig. \ref{f}f.

By changing the frequency of the loaded RF on the EOM, additional long-range coupling between different frequency modes can be established. A more complicated lattice structure~\cite{yuan2018synthetic} and the photonic gauge potential~\cite{dutt2019experimental} can be established. The dynamic modulation enables us to observe the dynamic bands~\cite{li2021dynamic}.
By using both frequency and amplitude modulation, the introduced gain and loss as well as long-range connections can be used to study arbitrary topological windings and braids of complex non-Hermitian bands~\cite{wang2021topological,wang2021generating}. Combining the synthetic frequency dimension with cavity arrays, one can study high-dimensional
~\cite{lin2016photonic,yuan2016photonic} and high-order~\cite{dutt2020higher} topological insulators. In the synthetic time-frequency space, a rich set of pulse propagation behaviors can be obtained~\cite{li2022single}. Even in a few fiber loops, one can obtain rich phenomena such as holographic quench dynamics~\cite{yu2021topological}, flat bands and band transitions~\cite{li2022observation}.
The boundary can also be created by coupling a cavity with different resonant frequencies~\cite{dutt2022creating}.
Moreover, the synthetic frequency dimension provides new designs for optical devices and computing devices, such as topologically protected mode-locked lasers~\cite{yang2020mode}, frequency domain mirrors~\cite{hu2022mirror}, and the multidimensional convolution processor~\cite{fan2022multidimensional}. \\

\noindent\textbf{Orbital angular momentum}\\
The first proposal of photonic synthetic dimensions is based on OAM~\cite{luo2015quantum,ozawa2019topological}. For synthetic OAM dimension, the photons carrying OAM $m\hbar$ have a phase term $e^{im\phi}$, where $m$ represents the integral topological charge, $\hbar$ represents the Planck constant, $\phi$ is the azimuthal angle. The OAM modes (denoted as $|m\rangle$) can be supported by a special cavity named degenerated optical cavity~\cite{cheng2017degenerate,cheng2018experimental,cheng2019flexible}, where all the OAM modes resonate at the same frequency. One of the possible constructions is shown in Fig. \ref{OAM}a. By introducing anisotropic liquid crystal molecules (Q-plate) into the cavity, the optical mode states $|L,m\rangle$ and $|R,m+1\rangle$ will couple with each other in coupling strength $J_{1}$, where $R$ $(L)$ represents the right (left)-circular polarization. Similarly, as one introduces another extra birefringent crystal (wave plate, WP) into the cavity, the optical mode states $|L,m\rangle$ and $|R,m\rangle$ will couple with coupling strength $J_{2}$. Thus we construct a Su-Schrieffer-Heeger (SSH) model along synthetic OAM dimension in a cavity, which is shown in Fig. \ref{OAM}b. As the coupling is weak, the Hamiltonian satisfies
\begin{equation}
\begin{aligned}
H_{O}=\sum_{m}(J_{1}\hat{a}_{R,m+1}^{\dag}\hat{a}_{L,m}+J_{2}\hat{a}_{R,m}^{\dag}\hat{a}_{L,m}+\mathrm{h.c.}).
\end{aligned}
\end{equation}
For the spatial symmetry of OAM lattice, we can rewrite the Hamiltonian in quasimomentum space ($k$-space) as $H_{O}=\int dk \hat{\bf a}_{k}^{\dag}H_{O}(k)\hat{\bf a}_{k}$, where $\hat{\bf a}_{k}=\sum_{m}e^{-imk}\hat{\bf a}_{m}$ and $\hat{\bf a}_{m}=(\hat{a}_{R,m}, \hat{a}_{L,m})^{T}$. The Hamiltonian in $k$-space can be expanded as $H_{O}(k)=\boldsymbol{\rm h}(k)\cdot\boldsymbol{\sigma}$, where $\boldsymbol{\sigma}=(\sigma_{x},\sigma_{y},\sigma_{z})$ are Pauli matrices and the eigen-energy $E_{O}(k)=|\boldsymbol{\rm h}(k)|=\sqrt{J_{1}^2+J_{2}^2+2J_{1}J_{2}\cos k}$. As an example, the calculated energy bands of $J_{1}=0.7$ and $J_{2}=0.3$ is shown in Fig. \ref{OAM}c. 

 \begin{figure*}[t!]
	\begin{center}
	\includegraphics[width=1.8 \columnwidth]{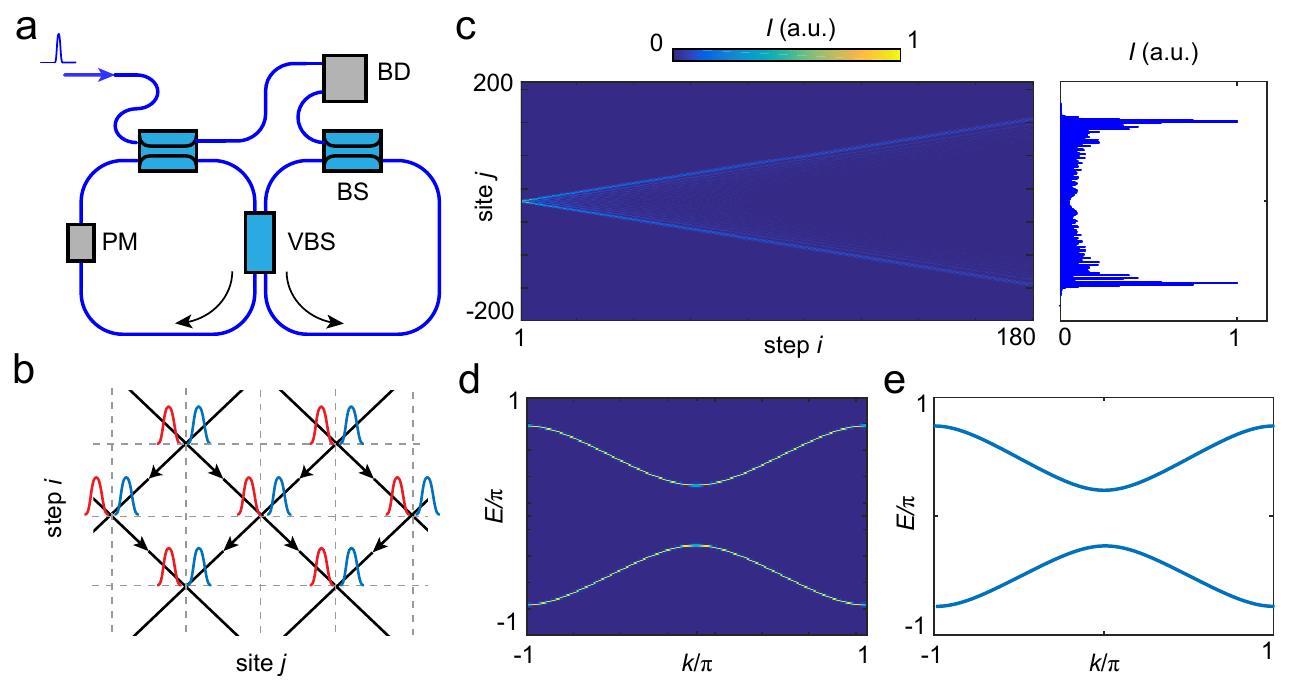}
		\caption{ \textbf{Synthetic time-multiplexed lattice.} (a) Simplified setup containing two coupled fiber loops. BS: beam spliter; PM: phase modulator; VBS: variable beam splitter; BD: balance detector; (b) The dynamics of the input pulse in a two dimensional lattice consisted of discrete time steps $i$ and site positions $j$. (c) Left: the graph of the spatio-temporal evolution of the injected pulse. Right: The pulse distribution as discrete time steps $i=180$. (d) Energy bands via two-dimensional Fourier transforming. (e) The calculated energy bands of the system.
		}
		\label{time}
	\end{center}
\end{figure*}

Similarly, the input and output relation along the synthetic OAM dimension is the same as the synthetic frequency dimension, and the transmitted coefficient satisfies Eq. \ref{T}. To detect the energy and OAM modes probability distribution information of this system, we can drive the cavity with the light carrying OAM $|m\rangle$, and excite the corresponding eigenstates of the cavity. The transmitted intensity spectrum vs frequency of incident photons as $J_{1}=0.7$, $J_{2}=0.3$ and $\gamma=0.05$ is shown in Fig. \ref{OAM}d, and no photon transmitting out of the cavity as the frequency located in the energy gap (white regions). Similarly, the transmitted intensity spectrum can be directly read out via an oscilloscope while linearly sweeping the frequency of the pumping light. 

Moreover, we can get the details of the energy bands by using wavefront angle-resolved band structure spectroscopy~\cite{yang2022realization}. We need to project the output photons on the base $|k\rangle\langle k|$, where $|k\rangle=\sum_{m}e^{-imk}|m\rangle$ is the Bloch state along OAM lattice. The wavefunction of OAM state can be represented by the complex amplitude, denoted as $|m\rangle=Ae^{im\phi}$, where $A$ is the amplitude. Thus the complex amplitude of Bloch state can be written as $|k\rangle=A\sum_{m}e^{-im(k-\phi)}=A\delta(k,\phi)$, which means the Bloch state corresponds to the photon state along azimuthal angle $\phi=k$ of the wavefront and is shown in the dashed pane in Fig. \ref{OAM}a. By detecting the transmitted intensity spectrum along different the azimuthal angle $\phi$, we can get the energy bands of the cavity as shown in Fig. \ref{OAM}e ($J_{1}=0.7$, $J_{2}=0.3$ and $\gamma=0.05$).
 
Through projective measurement of the OAM modes via forked grating~\cite{saitoh2013measuring} shown in Fig. \ref{OAM}a, we can get the energy population for different OAM modes when $J_{1}=0.7$, $J_{2}=0.3$ and $\gamma=0.05$ as shown in Fig. \ref{OAM}f. This results can be used to analyse quasiparticle distribution in synthetic lattices. Moreover, by introducing the projective measurement of polarizations, we would obtain the spin distribution information, which can be used to study spin-momentum locking and spin textures~\cite{guo2020meron}.

Rich physics would appear when modifying the structure of the cavity and engineering the Hamiltonian. We can investigate 2D topological physics
in a 1D array of optical cavities including edge-state transport and topological phase transition~\cite{luo2015quantum}. By drilling a hole on the wave plate, a sharp boundary can be created~\cite{zhou2017dynamically}. By introducing different loss on different polarizations, the exceptional points can be realised~\cite{yang2022realization}. Moreover, Weyl semimetal phases and implementation can be obtained along the synthetic OAM dimension in degenerate optical cavities~\cite{sun2017weyl}. Even in a single cavity, we can synthesize arbitrary lattice models~\cite{wang2019synthesizing}. The synthetic OAM dimension has been applied to form quantum memories, high order filters~\cite{luo2017synthetic} and OAM optical switchies~\cite{luo2018topological}. Due to the same framework of synthetic frequency and OAM dimensions, we can construct a two-dimensional lattice in gauge field in a single cavity with these two synthetic dimensions~\cite{yuan2019photonic}. \\



\noindent\textbf{Time-multiplexed lattice}\\
The synthetic lattice can also be formed by the photonic temporal degree of freedom. In two coupled fiber rings as shown in Fig. \ref{time}a, the pulse in one of the rings will be coupled into another ring after a round trip via the ariable beam splitter (VBS). The evolution of light can be mapped to a two dimensional discrete lattice along the round trip number (Fig. \ref{time}b), where red and blue pulses represent the pulses in left-moving and right-moving paths. The lattices are denoted by discrete time steps $i$, and the pulse site position $j$. Here we consider a coherent pulse of light coupled into the ring via the beam spliter (BS). The dynamics of the incident light pulse can be described by
\begin{equation}
\begin{aligned}
\alpha_{j}^{i+1}&=(\cos \theta_{i}\alpha_{j-1}^{i}+i\sin \theta_{i}\beta_{j-1}^{i})e^{i\varphi_{i}},\\
\beta_{j}^{i+1}&=i\sin \theta_{i}\alpha_{j-1}^{i}+\cos \theta_{i}\beta_{j-1}^{i},
\end{aligned}\label{walk}
\end{equation}
where $\alpha_{j}^{i}$ and $\beta_{j}^{i}$ represent the amplitude of the pulse in left-moving and right-moving paths, respectively. The parameter $\theta_{i}$ is controlled by the VBS at each time step $i$, where the transmission and reflection are $\sin\theta_{i}$ and $\cos\theta_{i}$, respectively. A phase modulator (PM) is used to change the phase $\varphi_{i}$ at a time step $i$. We can directly get the graph of spatiotemporal evolution by heterodyning the couple out light pulses with the input light pulse through a balanced detector (BD). 

For example, the graph of spatio-temporal evolution is shown in Fig. \ref{time}c left panel when $\theta_{i}=\pi/4$ and $\varphi_{i}=0$. The site distribution along pulse lattice at time step $i=180$ is shown in Fig. \ref{time}c right panel. Obviously, the quasi-particles (pulse modes) are mainly distributed at both sides, which is characteristic of a quantum walk. The graph of spatiotemporal evolution reveals the dynamics of the system in real space. The Bloch oscillation~\cite{regensburger2011photon} and Anderson localization~\cite{vatnik2017anderson} can be obtained in this synthetic system by flexibly changing the phase $\varphi_{i}$.
By coupling multiple fiber loops, a more complex network can be established~\cite{leefmans2022topological}.
Moreover, by introducing gain and loss into these two fiber loops, one can create a parity-time synthetic photonic lattice~\cite{regensburger2012parity}, where defect states~\cite{regensburger2013observation} and optical solitons~\cite{wimmer2015observation} have been obtained.
In a non-Hermitian temporal lattice, triple phase transitions~\cite{weidemann2022topological}, constant-intensity waves and induced transparency~\cite{steinfurth2022observation} have been investigated.
By engineering the skin effect and edges along the temporal lattice, one can synthesize a topological light funnel~\cite{weidemann2020topological}. In addition, this system can also be used to study physics in disorder~\cite{dikopoltsev2022observation}, and time crystals~\cite{taheri2022all}. 

On the other hand, for the translation symmetry and time periodicity, the pulse modes in the two rings can be rewritten as $(\alpha_{j}^{i}, \beta_{j}^{i})^{T}=(A,B)^{T}e^{-iE_{t}i}e^{-ikj}$, where $E_{t}$ is the quasi-energy. $A$ and $B$ are the amplitudes of the left-moving and right-moving optical pulses. The band structure can be obtained simply by applying the two-dimensional Fourier transform (2DFT) of the graph of spatio-temporal evolution~\cite{lechevalier2021single}. The 2DFT of Fig. \ref{time}c left panel is shown in Fig. \ref{time}d, which agrees well with the theoretical energy bands $\cos E=\cos k/\sqrt{2}$ in Fig. \ref{time}e. 2DFT of the graph of spatio-temporal evolution provides the sight of energy bands to investigate the topological phenomena, such as Berry curvature~\cite{wimmer2017experimental}, topological edge states~\cite{adiyatullin2022multi}, and Hofstadter butterfly~\cite{weidemann2022topological}.\\

\begin{figure}[t!]
	\begin{center}
	\includegraphics[width=0.9 \columnwidth]{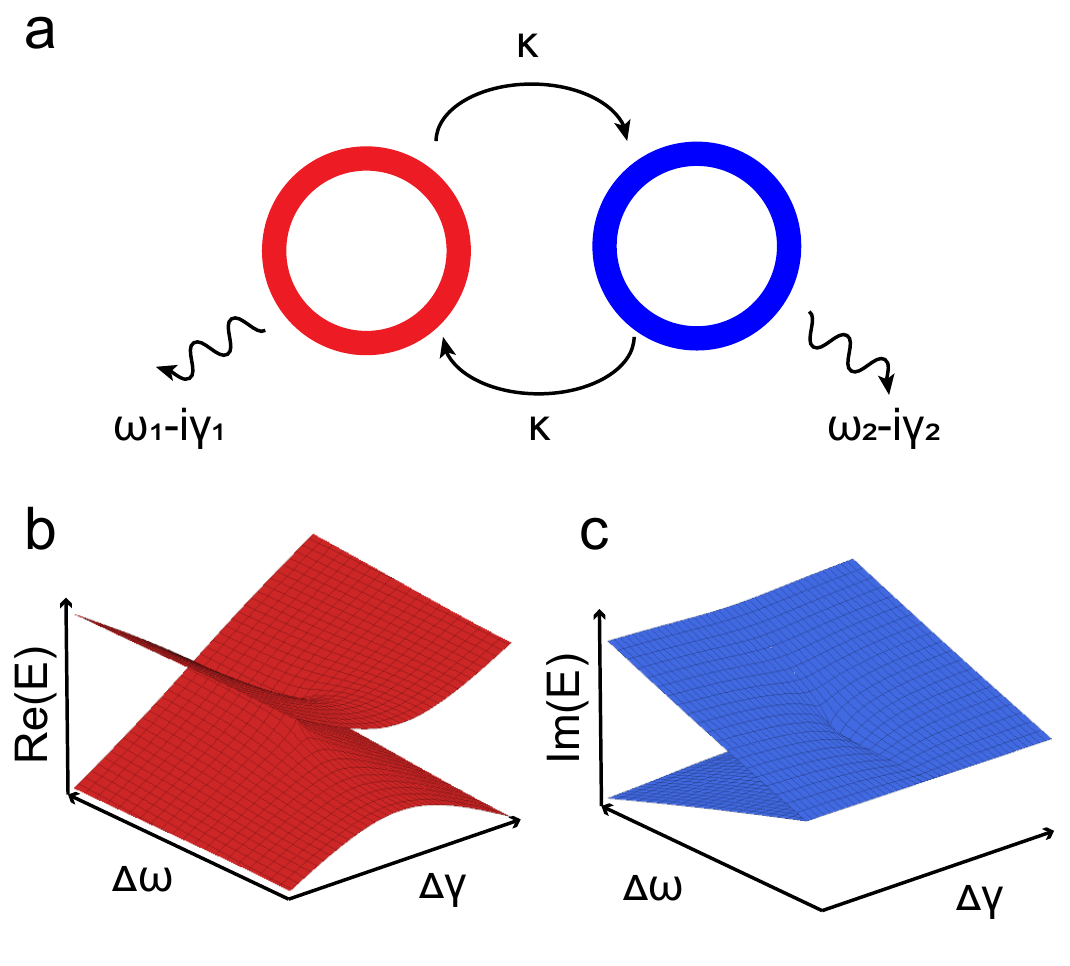}
		\caption{ \textbf{Synthetic parameter dimension.} (a) Two coupled microcavities with multiple independent parameters. (b) The real part of the energy in the parameter space. (c) The imaginary part of the energy in the parameter space.  
		}
		\label{para}
	\end{center}
\end{figure}

\noindent\textbf{Independent parameters}\\
The independent parameters in the cavities can usually be regarded as pseudo-momenta of the synthetic dimensions. To demonstrate this point, here we consider two coupled microcavities with coupling strength $\kappa$, as shown in Fig. \ref{para}a. Left (red) and right (blue) cavities have a resonance frequency of $\omega_{1}$ and $\omega_{2}$, respectively, and also have a decay rate of $\gamma_{1}$ and $\gamma_{2}$.
The amplitudes $a_{1}$ and $a_{2}$ in these cavities satisfy the coupled mode equations of time $t$, denoted as 
\begin{equation}
\frac{d}{dt}\left(
\begin{array}{c}
a_{1} \\
a_{2}
\end{array}
\right)
=-i
\left(
\begin{array}{cc}
\omega_{1}-i\gamma_{1}  & \kappa  \\
\kappa & \omega_{2}-i\gamma_{2}
\end{array}
\right)
\left(
\begin{array}{c}
a_{1} \\
a_{2}
\end{array}
\right).
\label{coupling}
\end{equation}
The solution of the amplitudes satisfies $(a_{1},a_{2})^{T}=(A_{1},A_{2})^{T} e^{-iE_{p}t}$. The eigen-energy of the system are
$E_{p}=\bar{\omega}-i\bar{\gamma}\pm\sqrt{\kappa^2+(\Delta\omega+i\Delta\gamma)^2}$, where $\bar{\omega}(\bar{\gamma})=(\omega_{1}(\gamma_{1})+\omega(\gamma)_{2})/2$, $\Delta\omega(\Delta\gamma)=(\omega_{1}(\gamma_{1})-\omega_{2}(\gamma_{2}))/2$ and $\Delta\gamma=(\gamma_{1}-\gamma_{2})/2$. The real and imaginary parts of the eigen-energy in the parameter space ($\Delta\omega, \Delta\gamma$) are shown in Fig. \ref{para}b and Fig. \ref{para}c, respectively. Interestingly, an exceptional point will occur at the point of $\Delta\omega=0$ and $\Delta\gamma=\pm\kappa$. Near the exceptional points, this system has important implications in the optical response for the singular topology. 

Large optical structures and devices have a similar principle, such as coupled cavities~\cite{peng2014parity,chang2014parity}, coupled optical and mechanical modes in an optomechanical cavity~\cite{verhagen2017optomechanical,xu2016topological,patil2022measuring}, and cavities within two-level atoms~\cite{lee2014heralded}. 
Independent parameter dimensions have been used to enhancing precision in sensing instruments~\cite{chen2017exceptional,hodaei2017enhanced,hodaei2014parity,langbein2018no}, multimode laser cavities~\cite{feng2014single}, and high sensitivity gyroscopes~\cite{lai2019observation}.
Moreover, combined with other synthetic dimensions, one can achieve higher physical dimensions, e. g., a two-dimensional array coupled with two independent parameters can achieve a four-dimensional quantum Hall effect~\cite{zilberberg2018photonic}. Although the independent parameters can be conveniently introduced in experiments, this synthetic dimension is parametric and is difficult to simulate the wavepacket transport in real space. \\

\noindent\textbf{Conclusion and prospects}\\
In this Review, we introduce simplified setups for constructing different photonic synthetic platforms. Their distinct characteristics and similarities have been characterized and discussed. With the extension of the simplified models, vast unique phenomena of topological materials have been simulated with photonic synthetic dimensions in cavities.

One of the prospects in synthetic dimensions is to investigate the many-body phase of matter by introducing photon-photon interactions in synthetic dimensions, since many-body physics is usually numerically difficult to study. Although the interaction among optical modes has been experimentally demonstrated, the interparticle (photon-photon) interactions along the photonic synthetic lattice are still experimentally challenge. A possible solution is to introduce light–matter (such as atoms) coupling in the cavity, known as mediate photonic interactions~\cite{chang2014quantum,birnbaum2005photon}. It may give birth to the intersection of synthetic dimensions and cavity quantum electrodynamics. For example, the photon-photon interaction can be achieved by Rydberg-mediated interactions. By Floquet engineering a twisted cavity, the photons are converted into strongly interacting polaritons~\cite{clark2019interacting}. The  Laughlin state has also been obtained in this configuration~\cite{clark2020observation}, which may open the way to observe two dimensional fractional quantum Hall phases. Moreover, the bound state can be investigated in giant atom-modulated resonators~\cite{xiao2022bound}.

On the other hand, the interphoton interaction can also be introduced by optical nonlinearity. For example, four-wave-mixing (FWM) processes can create the interacting Hamiltonian for the optical modes~\cite{yuan2020creating}, demoted as
$H_{\rm FWM}=\sum_{m,n,p,q}\hat{a}_{n}^{\dag}\hat{a}_{m}^{\dag}\hat{a}_{p}\hat{a}_{q},$
where $n+m=p+q$ satisfies the phase matching condition. The photon-blockade effect can be obtained by detecting the two-photon correlation probabilities. Moreover, optical solitons exist in the synthetic frequency lattice with an interacting Hamiltonian item~\cite{tusnin2020nonlinear,moille2022synthetic}. Notably, some classical optical detection might be ineffective, and more quantum detection methods should be explored, such as the detection of the two-photon correlation function and photon number resolving detection for quantized photonic synthetic dimensions. More importantly, studying the quantum behaviors of interphoton interactions usually require extremely low loss of the cavities. 

Another development direction is to integrate multiple photonic synthetic dimensions into a cavity and visualize high-dimensional topological matter with the optical detection methods. Photonic synthetic dimensions would enable us to observe the topological phase transitions by simply engineering the high-dimensional Hamiltonian. 
With the development of optical technologies, we expect the photonic synthesis dimensions to take optical computing advantages and quantum advantages compared with traditional von-Neumann computers. The photonic synthesis dimensions also show the potential to draw forth new optical devices, which would benefit optical and photonic information processing.\\

~\\
\noindent{\bf \large Acknowledgement}\\
This work was supported by the Innovation Program for Quantum Science and Technology (Grants No. 2021ZD0301400), the National Natural Science Foundation of China (Grants No. 61725504,  No. 11821404, No. U19A2075), the Anhui Initiative in Quantum Information Technologies (Grants No. AHY060300), the Fundamental Research Funds for the Central Universities (Grant No. WK2030380017, WK5290000003).

~\\
\noindent{\bf \large Additional Information}\\
The authors declare no competing financial interests.

~\\
\noindent{\bf \large Data Availability}\\
All of the data in this Review are available from the corresponding authors upon reasonable request.


\begin{thebibliography}{100}
 \bibitem{luo2015quantum}
 X.-W. Luo, X.~Zhou, C.-F. Li, J.-S. Xu, G.-C. Guo, and Z.-W. Zhou, ``Quantum
 simulation of 2d topological physics in a 1d array of optical cavities,''
 \emph{Nature communications}, vol.~6, no.~1, pp. 1--8, 2015.
 
 \bibitem{yuan2018synthetic}
 L.~Yuan, Q.~Lin, M.~Xiao, and S.~Fan, ``Synthetic dimension in photonics,''
 \emph{Optica}, vol.~5, no.~11, pp. 1396--1405, 2018.
 
 \bibitem{ozawa2019topological}
 T.~Ozawa and H.~M. Price, ``Topological quantum matter in synthetic
 dimensions,'' \emph{Nature Reviews Physics}, vol.~1, no.~5, pp. 349--357,
 2019.
 
 \bibitem{price2022roadmap}
 H.~Price, Y.~Chong, A.~Khanikaev, H.~Schomerus, L.~J. Maczewsky, M.~Kremer,
 M.~Heinrich, A.~Szameit, O.~Zilberberg, Y.~Yang \emph{et~al.}, ``Roadmap on
 topological photonics,'' \emph{Journal of Physics: Photonics}, 2022.
 
 \bibitem{mukherjee2020observation}
 S.~Mukherjee and M.~C. Rechtsman, ``Observation of floquet solitons in a
 topological bandgap,'' \emph{Science}, vol. 368, no. 6493, pp. 856--859,
 2020.
 
 \bibitem{maczewsky2020nonlinearity}
 L.~J. Maczewsky, M.~Heinrich, M.~Kremer, S.~K. Ivanov, M.~Ehrhardt,
 F.~Martinez, Y.~V. Kartashov, V.~V. Konotop, L.~Torner, D.~Bauer
 \emph{et~al.}, ``Nonlinearity-induced photonic topological insulator,''
 \emph{Science}, vol. 370, no. 6517, pp. 701--704, 2020.
 
 \bibitem{zhou2018observation}
 H.~Zhou, C.~Peng, Y.~Yoon, C.~W. Hsu, K.~A. Nelson, L.~Fu, J.~D. Joannopoulos,
 M.~Solja{\v{c}}i{\'c}, and B.~Zhen, ``Observation of bulk fermi arc and
 polarization half charge from paired exceptional points,'' \emph{Science},
 vol. 359, no. 6379, pp. 1009--1012, 2018.
 
 \bibitem{jin2019topologically}
 J.~Jin, X.~Yin, L.~Ni, M.~Solja{\v{c}}i{\'c}, B.~Zhen, and C.~Peng,
 ``Topologically enabled ultrahigh-q guided resonances robust to out-of-plane
 scattering,'' \emph{Nature}, vol. 574, no. 7779, pp. 501--504, 2019.
 
 \bibitem{yin2020observation}
 X.~Yin, J.~Jin, M.~Solja{\v{c}}i{\'c}, C.~Peng, and B.~Zhen, ``Observation of
 topologically enabled unidirectional guided resonances,'' \emph{Nature}, vol.
 580, no. 7804, pp. 467--471, 2020.
 
 \bibitem{stutzer2018photonic}
 S.~St{\"u}tzer, Y.~Plotnik, Y.~Lumer, P.~Titum, N.~H. Lindner, M.~Segev, M.~C.
 Rechtsman, and A.~Szameit, ``Photonic topological anderson insulators,''
 \emph{Nature}, vol. 560, no. 7719, pp. 461--465, 2018.
 
 \bibitem{rechtsman2013photonic}
 M.~C. Rechtsman, J.~M. Zeuner, Y.~Plotnik, Y.~Lumer, D.~Podolsky, F.~Dreisow,
 S.~Nolte, M.~Segev, and A.~Szameit, ``Photonic floquet topological
 insulators,'' \emph{Nature}, vol. 496, no. 7444, pp. 196--200, 2013.
 
 \bibitem{weimann2017topologically}
 S.~Weimann, M.~Kremer, Y.~Plotnik, Y.~Lumer, S.~Nolte, K.~G. Makris, M.~Segev,
 M.~C. Rechtsman, and A.~Szameit, ``Topologically protected bound states in
 photonic parity--time-symmetric crystals,'' \emph{Nature materials}, vol.~16,
 no.~4, pp. 433--438, 2017.
 
 \bibitem{bandres2018topological}
 M.~A. Bandres, S.~Wittek, G.~Harari, M.~Parto, J.~Ren, M.~Segev, D.~N.
 Christodoulides, and M.~Khajavikhan, ``Topological insulator laser:
 Experiments,'' \emph{Science}, vol. 359, no. 6381, p. eaar4005, 2018.
 
 \bibitem{hafezi2013imaging}
 M.~Hafezi, S.~Mittal, J.~Fan, A.~Migdall, and J.~Taylor, ``Imaging topological
 edge states in silicon photonics,'' \emph{Nature Photonics}, vol.~7, no.~12,
 pp. 1001--1005, 2013.
 
 \bibitem{lu2016topological}
 L.~Lu, J.~D. Joannopoulos, and M.~Solja{\v{c}}i{\'c}, ``Topological states in
 photonic systems,'' \emph{Nature Physics}, vol.~12, no.~7, pp. 626--629,
 2016.
 
 \bibitem{mittal2016measurement}
 S.~Mittal, S.~Ganeshan, J.~Fan, A.~Vaezi, and M.~Hafezi, ``Measurement of
 topological invariants in a 2d photonic system,'' \emph{Nature Photonics},
 vol.~10, no.~3, pp. 180--183, 2016.
 
 \bibitem{luo2017synthetic}
 X.-W. Luo, X.~Zhou, J.-S. Xu, C.-F. Li, G.-C. Guo, C.~Zhang, and Z.-W. Zhou,
 ``Synthetic-lattice enabled all-optical devices based on orbital angular
 momentum of light,'' \emph{Nature communications}, vol.~8, no.~1, pp. 1--7,
 2017.
 
 \bibitem{xiao2017observation}
 L.~Xiao, X.~Zhan, Z.~Bian, K.~Wang, X.~Zhang, X.~Wang, J.~Li, K.~Mochizuki,
 D.~Kim, N.~Kawakami \emph{et~al.}, ``Observation of topological edge states
 in parity--time-symmetric quantum walks,'' \emph{Nature Physics}, vol.~13,
 no.~11, pp. 1117--1123, 2017.
 
 \bibitem{xiao2020non}
 L.~Xiao, T.~Deng, K.~Wang, G.~Zhu, Z.~Wang, W.~Yi, and P.~Xue, ``Non-hermitian
 bulk--boundary correspondence in quantum dynamics,'' \emph{Nature Physics},
 vol.~16, no.~7, pp. 761--766, 2020.
 
 \bibitem{zhan2017detecting}
 X.~Zhan, L.~Xiao, Z.~Bian, K.~Wang, X.~Qiu, B.~C. Sanders, W.~Yi, and P.~Xue,
 ``Detecting topological invariants in nonunitary discrete-time quantum
 walks,'' \emph{Physical review letters}, vol. 119, no.~13, p. 130501, 2017.
 
 \bibitem{wang2019simulating}
 K.~Wang, X.~Qiu, L.~Xiao, X.~Zhan, Z.~Bian, W.~Yi, and P.~Xue, ``Simulating
 dynamic quantum phase transitions in photonic quantum walks,'' \emph{Physical
 	Review Letters}, vol. 122, no.~2, p. 020501, 2019.
 
 \bibitem{cardano2017detection}
 F.~Cardano, A.~D’Errico, A.~Dauphin, M.~Maffei, B.~Piccirillo, C.~de~Lisio,
 G.~De~Filippis, V.~Cataudella, E.~Santamato, L.~Marrucci \emph{et~al.},
 ``Detection of zak phases and topological invariants in a chiral quantum walk
 of twisted photons,'' \emph{Nature communications}, vol.~8, no.~1, pp. 1--7,
 2017.
 
 \bibitem{cardano2015quantum}
 F.~Cardano, F.~Massa, H.~Qassim, E.~Karimi, S.~Slussarenko, D.~Paparo,
 C.~de~Lisio, F.~Sciarrino, E.~Santamato, R.~W. Boyd \emph{et~al.}, ``Quantum
 walks and wavepacket dynamics on a lattice with twisted photons,''
 \emph{Science advances}, vol.~1, no.~2, p. e1500087, 2015.
 
 \bibitem{wang2018dynamic}
 Q.-Q. Wang, X.-Y. Xu, W.-W. Pan, K.~Sun, J.-S. Xu, G.~Chen, Y.-J. Han, C.-F.
 Li, and G.-C. Guo, ``Dynamic-disorder-induced enhancement of entanglement in
 photonic quantum walks,'' \emph{Optica}, vol.~5, no.~9, pp. 1136--1140, 2018.
 
 \bibitem{xu2018measuring}
 X.-Y. Xu, Q.-Q. Wang, W.-W. Pan, K.~Sun, J.-S. Xu, G.~Chen, J.-S. Tang,
 M.~Gong, Y.-J. Han, C.-F. Li \emph{et~al.}, ``Measuring the winding number in
 a large-scale chiral quantum walk,'' \emph{Physical Review Letters}, vol.
 120, no.~26, p. 260501, 2018.
 
 \bibitem{yuan2021synthetic}
 L.~Yuan, A.~Dutt, and S.~Fan, ``Synthetic frequency dimensions in dynamically
 modulated ring resonators,'' \emph{APL Photonics}, vol.~6, no.~7, p. 071102,
 2021.
 
 \bibitem{leykam2020topological}
 D.~Leykam and L.~Yuan, ``Topological phases in ring resonators: recent progress
 and future prospects,'' \emph{Nanophotonics}, vol.~9, no.~15, pp. 4473--4487,
 2020.
 
 \bibitem{yang2022topological}
 M.~Yang, H.-Q. Zhang, Y.-W. Liao, Z.-H. Liu, Z.-W. Zhou, X.-X. Zhou, J.-S. Xu,
 Y.-J. Han, C.-F. Li, and G.-C. Guo, ``Topological band structure via twisted
 photons in a degenerate cavity,'' \emph{Nature communications}, vol.~13,
 no.~1, pp. 1--7, 2022.
 
 \bibitem{regensburger2012parity}
 A.~Regensburger, C.~Bersch, M.-A. Miri, G.~Onishchukov, D.~N. Christodoulides,
 and U.~Peschel, ``Parity-time synthetic photonic lattices,'' \emph{Nature},
 vol. 488, no. 7410, pp. 167--171, 2012.
 
 \bibitem{adiyatullin2022multi}
 A.~F. Adiyatullin, L.~K. Upreti, C.~Lechevalier, C.~Evain, F.~Copie, P.~Suret,
 S.~Randoux, P.~Delplace, and A.~Amo, ``Multi-topological floquet metals in a
 photonic lattice,'' \emph{arXiv preprint arXiv:2203.01056}, 2022.
 
 \bibitem{miri2019exceptional}
 M.-A. Miri and A.~Al{\`u}, ``Exceptional points in optics and photonics,''
 \emph{Science}, vol. 363, no. 6422, p. eaar7709, 2019.
 
 \bibitem{yuan2016bloch}
 L.~Yuan and S.~Fan, ``Bloch oscillation and unidirectional translation of
 frequency in a dynamically modulated ring resonator,'' \emph{Optica}, vol.~3,
 no.~9, pp. 1014--1018, 2016.
 
 \bibitem{scully1999quantum}
 M.~O. Scully and M.~S. Zubairy, ``Quantum optics,'' 1999.
 
 \bibitem{li2021dynamic}
 G.~Li, Y.~Zheng, A.~Dutt, D.~Yu, Q.~Shan, S.~Liu, L.~Yuan, S.~Fan, and X.~Chen,
 ``Dynamic band structure measurement in the synthetic space,'' \emph{Science
 	advances}, vol.~7, no.~2, p. eabe4335, 2021.
 
 \bibitem{dutt2020single}
 A.~Dutt, Q.~Lin, L.~Yuan, M.~Minkov, M.~Xiao, and S.~Fan, ``A single photonic
 cavity with two independent physical synthetic dimensions,'' \emph{Science},
 vol. 367, no. 6473, pp. 59--64, 2020.
 
 \bibitem{dutt2019experimental}
 A.~Dutt, M.~Minkov, Q.~Lin, L.~Yuan, D.~A. Miller, and S.~Fan, ``Experimental
 band structure spectroscopy along a synthetic dimension,'' \emph{Nature
 	communications}, vol.~10, no.~1, pp. 1--8, 2019.
 
 \bibitem{wang2021topological}
 K.~Wang, A.~Dutt, C.~C. Wojcik, and S.~Fan, ``Topological complex-energy
 braiding of non-hermitian bands,'' \emph{Nature}, vol. 598, no. 7879, pp.
 59--64, 2021.
 
 \bibitem{wang2021generating}
 K.~Wang, A.~Dutt, K.~Y. Yang, C.~C. Wojcik, J.~Vu{\v{c}}kovi{\'c}, and S.~Fan,
 ``Generating arbitrary topological windings of a non-hermitian band,''
 \emph{Science}, vol. 371, no. 6535, pp. 1240--1245, 2021.
 
 \bibitem{lin2016photonic}
 Q.~Lin, M.~Xiao, L.~Yuan, and S.~Fan, ``Photonic weyl point in a
 two-dimensional resonator lattice with a synthetic frequency dimension,''
 \emph{Nature communications}, vol.~7, no.~1, pp. 1--7, 2016.
 
 \bibitem{yuan2016photonic}
 L.~Yuan, Y.~Shi, and S.~Fan, ``Photonic gauge potential in a system with a
 synthetic frequency dimension,'' \emph{Optics letters}, vol.~41, no.~4, pp.
 741--744, 2016.
 
 \bibitem{dutt2020higher}
 A.~Dutt, M.~Minkov, I.~A. Williamson, and S.~Fan, ``Higher-order topological
 insulators in synthetic dimensions,'' \emph{Light: Science \& Applications},
 vol.~9, no.~1, pp. 1--9, 2020.
 
 \bibitem{li2022single}
 G.~Li, D.~Yu, L.~Yuan, and X.~Chen, ``Single pulse manipulations in synthetic
 time-frequency space,'' \emph{Laser \& Photonics Reviews}, vol.~16, no.~1, p.
 2100340, 2022.
 
 \bibitem{yu2021topological}
 D.~Yu, B.~Peng, X.~Chen, X.-J. Liu, and L.~Yuan, ``Topological holographic
 quench dynamics in a synthetic frequency dimension,'' \emph{Light: Science \&
 	Applications}, vol.~10, no.~1, pp. 1--11, 2021.
 
 \bibitem{li2022observation}
 G.~Li, L.~Wang, R.~Ye, S.~Liu, Y.~Zheng, L.~Yuan, and X.~Chen, ``Observation of
 flat-band and band transition in the synthetic space,'' \emph{Advanced
 	Photonics}, vol.~4, no.~3, p. 036002, 2022.
 
 \bibitem{dutt2022creating}
 A.~Dutt, L.~Yuan, K.~Y. Yang, K.~Wang, S.~Buddhiraju, J.~Vu{\v{c}}kovi{\'c},
 and S.~Fan, ``Creating boundaries along a synthetic frequency dimension,''
 \emph{Nature Communications}, vol.~13, no.~1, p. 3377, 2022.
 
 \bibitem{yang2020mode}
 Z.~Yang, E.~Lustig, G.~Harari, Y.~Plotnik, Y.~Lumer, M.~A. Bandres, and
 M.~Segev, ``Mode-locked topological insulator laser utilizing synthetic
 dimensions,'' \emph{Physical Review X}, vol.~10, no.~1, p. 011059, 2020.
 
 \bibitem{hu2022mirror}
 Y.~Hu, M.~Yu, N.~Sinclair, D.~Zhu, R.~Cheng, C.~Wang, and M.~Lon{\v{c}}ar,
 ``Mirror-induced reflection in the frequency domain,'' \emph{Nature
 	communications}, vol.~13, no.~1, pp. 1--9, 2022.
 
 \bibitem{fan2022multidimensional}
 L.~Fan, Z.~Zhao, K.~Wang, A.~Dutt, J.~Wang, S.~Buddhiraju, C.~C. Wojcik, and
 S.~Fan, ``Multidimensional convolution operation with synthetic frequency
 dimensions in photonics,'' \emph{Physical Review Applied}, vol.~18, no.~3, p.
 034088, 2022.
 
 \bibitem{cheng2017degenerate}
 Z.-D. Cheng, Z.-D. Liu, X.-W. Luo, Z.-W. Zhou, J.~Wang, Q.~Li, Y.-T. Wang,
 J.-S. Tang, J.-S. Xu, C.-F. Li \emph{et~al.}, ``Degenerate cavity supporting
 more than 31 laguerre--gaussian modes,'' \emph{Optics letters}, vol.~42,
 no.~10, pp. 2042--2045, 2017.
 
 \bibitem{cheng2018experimental}
 Z.-D. Cheng, Q.~Li, Z.-H. Liu, F.-F. Yan, S.~Yu, J.-S. Tang, Z.-W. Zhou, J.-S.
 Xu, C.-F. Li, and G.-C. Guo, ``Experimental implementation of a degenerate
 optical resonator supporting more than 46 laguerre-gaussian modes,''
 \emph{Applied Physics Letters}, vol. 112, no.~20, p. 201104, 2018.
 
 \bibitem{cheng2019flexible}
 Z.-D. Cheng, Z.-H. Liu, Q.~Li, Z.-W. Zhou, J.-S. Xu, C.-F. Li, and G.-C. Guo,
 ``Flexible degenerate cavity with ellipsoidal mirrors,'' \emph{Optics
 	Letters}, vol.~44, no.~21, pp. 5254--5257, 2019.
 
 \bibitem{yang2022realization}
 M.~Yang, H.-Q. Zhang, Y.-W. Liao, Z.-H. Liu, Z.-W. Zhou, X.-X. Zhou, J.-S. Xu,
 Y.-J. Han, C.-F. Li, and G.-C. Guo, ``Realization of exceptional points along
 a synthetic orbital angular momentum dimension,'' \emph{arXiv preprint
 	arXiv:2209.07769}, 2022.
 
 \bibitem{saitoh2013measuring}
 K.~Saitoh, Y.~Hasegawa, K.~Hirakawa, N.~Tanaka, and M.~Uchida, ``Measuring the
 orbital angular momentum of electron vortex beams using a forked grating,''
 \emph{Physical review letters}, vol. 111, no.~7, p. 074801, 2013.
 
 \bibitem{guo2020meron}
 C.~Guo, M.~Xiao, Y.~Guo, L.~Yuan, and S.~Fan, ``Meron spin textures in momentum
 space,'' \emph{Physical review letters}, vol. 124, no.~10, p. 106103, 2020.
 
 \bibitem{zhou2017dynamically}
 X.-F. Zhou, X.-W. Luo, S.~Wang, G.-C. Guo, X.~Zhou, H.~Pu, and Z.-W. Zhou,
 ``Dynamically manipulating topological physics and edge modes in a single
 degenerate optical cavity,'' \emph{Physical review letters}, vol. 118, no.~8,
 p. 083603, 2017.
 
 \bibitem{sun2017weyl}
 B.~Y. Sun, X.~W. Luo, M.~Gong, G.~C. Guo, and Z.~W. Zhou, ``Weyl semimetal
 phases and implementation in degenerate optical cavities,'' \emph{Physical
 	Review A}, vol.~96, no.~1, p. 013857, 2017.
 
 \bibitem{wang2019synthesizing}
 S.~Wang, X.-F. Zhou, G.-C. Guo, H.~Pu, and Z.-W. Zhou, ``Synthesizing arbitrary
 lattice models using a single degenerate cavity,'' \emph{Physical Review A},
 vol. 100, no.~4, p. 043817, 2019.
 
 \bibitem{luo2018topological}
 X.-W. Luo, C.~Zhang, G.-C. Guo, and Z.-W. Zhou, ``Topological photonic
 orbital-angular-momentum switch,'' \emph{Physical Review A}, vol.~97, no.~4,
 p. 043841, 2018.
 
 \bibitem{yuan2019photonic}
 L.~Yuan, Q.~Lin, A.~Zhang, M.~Xiao, X.~Chen, and S.~Fan, ``Photonic gauge
 potential in one cavity with synthetic frequency and orbital angular momentum
 dimensions,'' \emph{Physical Review Letters}, vol. 122, no.~8, p. 083903,
 2019.
 
 \bibitem{regensburger2011photon}
 A.~Regensburger, C.~Bersch, B.~Hinrichs, G.~Onishchukov, A.~Schreiber,
 C.~Silberhorn, and U.~Peschel, ``Photon propagation in a discrete fiber
 network: An interplay of coherence and losses,'' \emph{Physical review
 	letters}, vol. 107, no.~23, p. 233902, 2011.
 
 \bibitem{vatnik2017anderson}
 I.~D. Vatnik, A.~Tikan, G.~Onishchukov, D.~V. Churkin, and A.~A. Sukhorukov,
 ``Anderson localization in synthetic photonic lattices,'' \emph{Scientific
 	reports}, vol.~7, no.~1, pp. 1--6, 2017.
 
 \bibitem{leefmans2022topological}
 C.~Leefmans, A.~Dutt, J.~Williams, L.~Yuan, M.~Parto, F.~Nori, S.~Fan, and
 A.~Marandi, ``Topological dissipation in a time-multiplexed photonic
 resonator network,'' \emph{Nature Physics}, vol.~18, no.~4, pp. 442--449,
 2022.
 
 \bibitem{regensburger2013observation}
 A.~Regensburger, M.-A. Miri, C.~Bersch, J.~N{\"a}ger, G.~Onishchukov, D.~N.
 Christodoulides, and U.~Peschel, ``Observation of defect states in
 pt-symmetric optical lattices,'' \emph{Physical review letters}, vol. 110,
 no.~22, p. 223902, 2013.
 
 \bibitem{wimmer2015observation}
 M.~Wimmer, A.~Regensburger, M.-A. Miri, C.~Bersch, D.~N. Christodoulides, and
 U.~Peschel, ``Observation of optical solitons in pt-symmetric lattices,''
 \emph{Nature communications}, vol.~6, no.~1, pp. 1--9, 2015.
 
 \bibitem{weidemann2022topological}
 S.~Weidemann, M.~Kremer, S.~Longhi, and A.~Szameit, ``Topological triple phase
 transition in non-hermitian floquet quasicrystals,'' \emph{Nature}, vol. 601,
 no. 7893, pp. 354--359, 2022.
 
 \bibitem{steinfurth2022observation}
 A.~Steinfurth, I.~Kre{\v{s}}i{\'c}, S.~Weidemann, M.~Kremer, K.~G. Makris,
 M.~Heinrich, S.~Rotter, and A.~Szameit, ``Observation of photonic
 constant-intensity waves and induced transparency in tailored non-hermitian
 lattices,'' \emph{Science Advances}, vol.~8, no.~21, p. eabl7412, 2022.
 
 \bibitem{weidemann2020topological}
 S.~Weidemann, M.~Kremer, T.~Helbig, T.~Hofmann, A.~Stegmaier, M.~Greiter,
 R.~Thomale, and A.~Szameit, ``Topological funneling of light,''
 \emph{Science}, vol. 368, no. 6488, pp. 311--314, 2020.
 
 \bibitem{dikopoltsev2022observation}
 A.~Dikopoltsev, S.~Weidemann, M.~Kremer, A.~Steinfurth, H.~H. Sheinfux,
 A.~Szameit, and M.~Segev, ``Observation of anderson localization beyond the
 spectrum of the disorder,'' \emph{Science Advances}, vol.~8, no.~21, p.
 eabn7769, 2022.
 
 \bibitem{taheri2022all}
 H.~Taheri, A.~B. Matsko, L.~Maleki, and K.~Sacha, ``All-optical dissipative
 discrete time crystals,'' \emph{Nature communications}, vol.~13, no.~1, pp.
 1--10, 2022.
 
 \bibitem{lechevalier2021single}
 C.~Lechevalier, C.~Evain, P.~Suret, F.~Copie, A.~Amo, and S.~Randoux,
 ``Single-shot measurement of the photonic band structure in a fiber-based
 floquet-bloch lattice,'' \emph{Communications Physics}, vol.~4, no.~1, pp.
 1--9, 2021.
 
 \bibitem{wimmer2017experimental}
 M.~Wimmer, H.~M. Price, I.~Carusotto, and U.~Peschel, ``Experimental
 measurement of the berry curvature from anomalous transport,'' \emph{Nature
 	Physics}, vol.~13, no.~6, pp. 545--550, 2017.
 
 \bibitem{peng2014parity}
 B.~Peng, {\c{S}}.~K. {\"O}zdemir, F.~Lei, F.~Monifi, M.~Gianfreda, G.~L. Long,
 S.~Fan, F.~Nori, C.~M. Bender, and L.~Yang, ``Parity--time-symmetric
 whispering-gallery microcavities,'' \emph{Nature Physics}, vol.~10, no.~5,
 pp. 394--398, 2014.
 
 \bibitem{chang2014parity}
 L.~Chang, X.~Jiang, S.~Hua, C.~Yang, J.~Wen, L.~Jiang, G.~Li, G.~Wang, and
 M.~Xiao, ``Parity--time symmetry and variable optical isolation in
 active--passive-coupled microresonators,'' \emph{Nature photonics}, vol.~8,
 no.~7, pp. 524--529, 2014.
 
 \bibitem{verhagen2017optomechanical}
 E.~Verhagen and A.~Al{\`u}, ``Optomechanical nonreciprocity,'' \emph{Nature
 	Physics}, vol.~13, no.~10, pp. 922--924, 2017.
 
 \bibitem{xu2016topological}
 H.~Xu, D.~Mason, L.~Jiang, and J.~Harris, ``Topological energy transfer in an
 optomechanical system with exceptional points,'' \emph{Nature}, vol. 537, no.
 7618, pp. 80--83, 2016.
 
 \bibitem{patil2022measuring}
 Y.~S. Patil, J.~H{\"o}ller, P.~A. Henry, C.~Guria, Y.~Zhang, L.~Jiang,
 N.~Kralj, N.~Read, and J.~G. Harris, ``Measuring the knot of non-hermitian
 degeneracies and non-commuting braids,'' \emph{Nature}, vol. 607, no. 7918,
 pp. 271--275, 2022.
 
 \bibitem{lee2014heralded}
 T.~E. Lee and C.-K. Chan, ``Heralded magnetism in non-hermitian atomic
 systems,'' \emph{Physical Review X}, vol.~4, no.~4, p. 041001, 2014.
 
 \bibitem{chen2017exceptional}
 W.~Chen, {\c{S}}.~Kaya~{\"O}zdemir, G.~Zhao, J.~Wiersig, and L.~Yang,
 ``Exceptional points enhance sensing in an optical microcavity,''
 \emph{Nature}, vol. 548, no. 7666, pp. 192--196, 2017.
 
 \bibitem{hodaei2017enhanced}
 H.~Hodaei, A.~U. Hassan, S.~Wittek, H.~Garcia-Gracia, R.~El-Ganainy, D.~N.
 Christodoulides, and M.~Khajavikhan, ``Enhanced sensitivity at higher-order
 exceptional points,'' \emph{Nature}, vol. 548, no. 7666, pp. 187--191, 2017.
 
 \bibitem{hodaei2014parity}
 H.~Hodaei, M.-A. Miri, M.~Heinrich, D.~N. Christodoulides, and M.~Khajavikhan,
 ``Parity-time--symmetric microring lasers,'' \emph{Science}, vol. 346, no.
 6212, pp. 975--978, 2014.
 
 \bibitem{langbein2018no}
 W.~Langbein, ``No exceptional precision of exceptional-point sensors,''
 \emph{Physical Review A}, vol.~98, no.~2, p. 023805, 2018.
 
 \bibitem{feng2014single}
 L.~Feng, Z.~J. Wong, R.-M. Ma, Y.~Wang, and X.~Zhang, ``Single-mode laser by
 parity-time symmetry breaking,'' \emph{Science}, vol. 346, no. 6212, pp.
 972--975, 2014.
 
 \bibitem{lai2019observation}
 Y.-H. Lai, Y.-K. Lu, M.-G. Suh, Z.~Yuan, and K.~Vahala, ``Observation of the
 exceptional-point-enhanced sagnac effect,'' \emph{Nature}, vol. 576, no.
 7785, pp. 65--69, 2019.
 
 \bibitem{zilberberg2018photonic}
 O.~Zilberberg, S.~Huang, J.~Guglielmon, M.~Wang, K.~P. Chen, Y.~E. Kraus, and
 M.~C. Rechtsman, ``Photonic topological boundary pumping as a probe of 4d
 quantum hall physics,'' \emph{Nature}, vol. 553, no. 7686, pp. 59--62, 2018.
 
 \bibitem{chang2014quantum}
 D.~E. Chang, V.~Vuleti{\'c}, and M.~D. Lukin, ``Quantum nonlinear
 optics—photon by photon,'' \emph{Nature Photonics}, vol.~8, no.~9, pp.
 685--694, 2014.
 
 \bibitem{birnbaum2005photon}
 K.~M. Birnbaum, A.~Boca, R.~Miller, A.~D. Boozer, T.~E. Northup, and H.~J.
 Kimble, ``Photon blockade in an optical cavity with one trapped atom,''
 \emph{Nature}, vol. 436, no. 7047, pp. 87--90, 2005.
 
 \bibitem{clark2019interacting}
 L.~W. Clark, N.~Jia, N.~Schine, C.~Baum, A.~Georgakopoulos, and J.~Simon,
 ``Interacting floquet polaritons,'' \emph{Nature}, vol. 571, no. 7766, pp.
 532--536, 2019.
 
 \bibitem{clark2020observation}
 L.~W. Clark, N.~Schine, C.~Baum, N.~Jia, and J.~Simon, ``Observation of
 laughlin states made of light,'' \emph{Nature}, vol. 582, no. 7810, pp.
 41--45, 2020.
 
 \bibitem{xiao2022bound}
 H.~Xiao, L.~Wang, Z.-H. Li, X.~Chen, and L.~Yuan, ``Bound state in a giant
 atom-modulated resonators system,'' \emph{npj Quantum Information}, vol.~8,
 no.~1, pp. 1--7, 2022.
 
 \bibitem{yuan2020creating}
 L.~Yuan, A.~Dutt, M.~Qin, S.~Fan, and X.~Chen, ``Creating locally interacting
 hamiltonians in the synthetic frequency dimension for photons,''
 \emph{Photonics Research}, vol.~8, no.~9, pp. B8--B14, 2020.
 
 \bibitem{tusnin2020nonlinear}
 A.~K. Tusnin, A.~M. Tikan, and T.~J. Kippenberg, ``Nonlinear states and
 dynamics in a synthetic frequency dimension,'' \emph{Physical Review A}, vol.
 102, no.~2, p. 023518, 2020.
 
 \bibitem{moille2022synthetic}
 G.~Moille, C.~Menyuk, Y.~K. Chembo, A.~Dutt, and K.~Srinivasan, ``Synthetic
 frequency lattices from an integrated dispersive multi-color soliton,''
 \emph{arXiv preprint arXiv:2210.09036}, 2022.
 
\end{thebibliography}

\end{document}